\documentclass{aa}
\usepackage{graphicx}
\usepackage{amsmath}
\RequirePackage{natbib}
\begin{document}
\title{Vorticity in the solar photosphere}
\titlerunning{Vorticity in the solar atmosphere}
\author{S.~Shelyag, P.~Keys, M.~Mathioudakis, F.P.~Keenan}
\authorrunning{S.~Shelyag et al.}
\institute{Astrophysics Research Centre, School of Mathematics and Physics, Queen's University Belfast, Belfast,
BT7 1NN, Northern Ireland, UK}

\date{01.01.01/01.01.01}

\abstract {}
{We use magnetic and non-magnetic 3D numerical simulations of solar granulation and G-band radiative diagnostics from the 
resulting models to analyse the generation of small-scale vortex motions in the solar photosphere.}
{Radiative MHD simulations of magnetoconvection are used to produce photospheric models. Our starting point 
is a non-magnetic model of solar convection, where we introduce a uniform magnetic field and follow the evolution of the 
field in the simulated photosphere. We find two different types of photospheric vortices, and provide a link 
between the vorticity generation and the presence of the intergranular magnetic field. A detailed analysis of the vorticity 
equation, combined with the G-band radiative diagnostics, allows us to identify the sources and observational signatures of 
photospheric vorticity in the simulated photosphere. }
{Two different types of photospheric vorticity, magnetic and non-magnetic, are generated in the domain. Non-magnetic 
vortices are generated by the baroclinic motions of the plasma in the photosphere, while magnetic vortices are produced by 
the magnetic tension in the intergranular magnetic flux concentrations. The two types of vortices have different shapes. We find that 
the vorticity is generated more efficiently in the magnetised model. Simulated G-band images show a direct connection 
between magnetic vortices and rotary motions of photospheric bright points, and suggest that there may be a connection between 
the magnetic bright point rotation and small-scale swirl motions observed higher in the atmosphere.
}
{}


\maketitle

\section{Introduction}

One of the consequences of turbulent plasma movement in the solar photosphere is the appearance of horizontal vortex motions. 
These motions can be of significance for the generation of magneto-hydrodynamic (MHD) waves which propagate to the 
upper layers of the solar atmosphere \citep{parker1,fedun1,Jess1}. As a result of recent advancements in state-of-the-art 
instrumentation and observational techniques, it has now become possible to observe small-scale vortices in the lower 
solar atmosphere. 

\citet{bonet1} showed the presence of vortex 
motions in photospheric G-band bright points with lifetimes comparable to those of the granules. \citet{wedemeyer1, wedemeyer2} 
performed simultaneous G-band and Ca\,{\sc ii} $8542\mathrm{\AA}$ imaging to demonstrate the presence of small-scale swirl motions 
in the chromosphere. \citet{carlsson2} also demonstrate a presence of the chromospheric swirls in their simulations, which include
the chromospheric layer.

A thorough investigation of the non-magnetic photospheric convection by \citet{stein1} has shown the generation of 
vorticity by baroclinic fluid motions in the upper convection zone. Such motions, characterised by large and non-parallel
gradients of density and pressure, occur near the edges of granules. Some studies have also been performed on the 
vorticity generated by magnetic flux tubes rising to the solar surface from the deep sub-photosphere  \citep[e.g.][]{emonet1,emonet2}.  
The link between vortex generation and magnetic field in the upper photosphere was also noted by \citet{voegler1}.   
Generation of Alfv{\'e}nic vortices by the interaction of compressible plasma with field-aligned obstacles was studied by \citet{nakariakov1}.
Recently, \citet{kosovichev1} used numerical simulations to demonstrate the significance of vortex motions 
and vortex dragging for the creation of pore-like magnetised structures in the photosphere. 

Here we present a more detailed study of the photospheric vorticity. We focus on its origins and connection to the photospheric
magnetic field and granulation dynamics. We show a direct correspondence between the photospheric vortices that 
correspond to G-band bright point motions, with the strong intergranular magnetic field. We also demonstrate a presence 
of vortex motions in the upper atmosphere, which may be connected to the chromospheric swirls.

In Section 2 we provide a brief description of the numerical model, simulation setup and the results obtained. 
The output of the simulations is analysed in terms of the vorticity equation in Section 3. Section 4 describes the
radiative diagnostics and observational consequences of vortex motions in the photosphere, while in Section 5  we summarise our conclusions. 

\begin{figure*}
\includegraphics[width=1.0\linewidth]{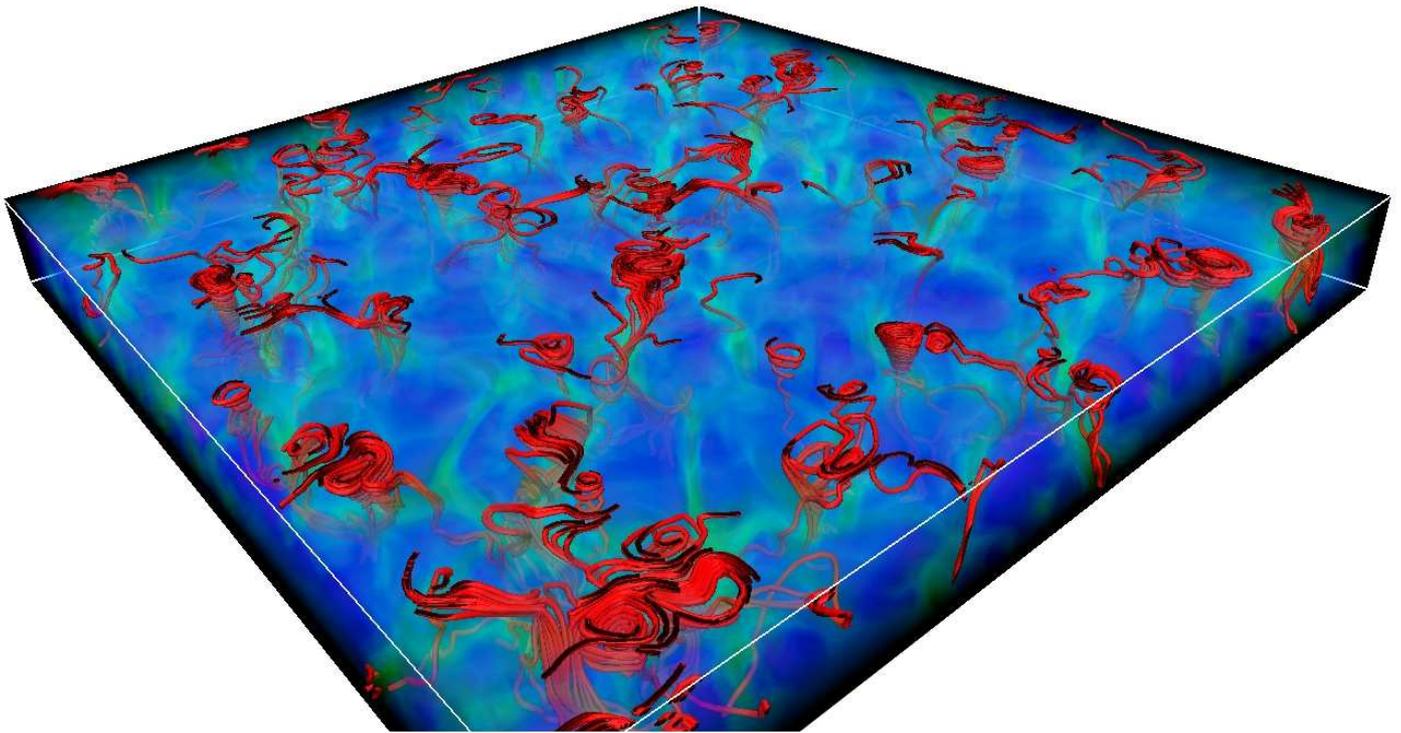}
\caption{Geometry of the domain. Stream lines (red) are plotted over the three-dimensional structure of the vertical 
component of the magnetic field (blue-green). The image is generated with the VAPOR 3D visualisation package \citep{vapor}.}
\label{fig0}
\end{figure*}

\section{Simulations}

We use the MuRAM code \citep{voegler1} to perform the simulations, which has been successfully used
for a wide range of solar applications \citep{shelyagbp1,shelyagbp2,shelyag2,cheung1,pietarila1,rempel1,yelles1,danilovic2}. 
The code solves large-eddy radiative three-dimensional MHD equations on a Cartesian grid, and employs a fourth-order 
central difference scheme to calculate spatial derivatives. A fourth-order Runge-Kutta scheme is used to advance the 
numerical solution in time. Hyperdiffusivity sources are used to stabilise the solution against numerical instabilities 
and to account for physical processes which are not resolved by the numerical grid. Non-grey radiative energy 
transport is included in the code using short-characteristics and opacity binning techniques. A non-ideal 
equation-of-state, taking into account the first-stage ionisation of the 11 most abundant elements in the Sun, is also included.

The numerical domain has a physical size of $12\times 12~\mathrm{Mm}^2$ in the horizontal direction,  
$1.4~\mathrm{Mm}$ in the vertical direction ($y$), and is resolved by $480\times480$ and 100 grid cells,
respectively. The lower boundary is transparent, allowing the plasma to travel in and out of the domain. The 
upper boundary of the domain is closed, however, it allows the horizontal motions of the plasma and the magnetic field lines. The side 
boundaries of the domain are set to be periodic. The domain is positioned in such a way that the visible solar surface\footnote{In 
this paper we refer to the visible solar surface as a horizontal geometrical layer which is physically close to the optical layer 
of radiation formation. We assume that the geometrical properties of the analysed features do not significantly change within this layer.} is located
approximately $600~\mathrm{km}$ below the upper boundary.

Our starting point for the simulations is a well-developed non-magnetic ($\mathbf{B}=0$) snapshot of photospheric convection taken 
at $t \sim 2000~\mathrm{s}$ (about 8 convective turnover timescales) from the initial plane-parallel model. A uniform vertical 
magnetic field of $B_y=200~\mathrm{G}$ has been introduced at this stage, and a sequence of 147 snapshots recorded, 
containing physical parameters of the model, such as velocity and magnetic field vectors, temperature, density, pressure and 
internal energy. The sequence covers approximately 40 minutes of physical time, corresponding to $\sim$5-10 granular lifetimes. 
During the simulation, the magnetic field is advected into the intergranular lanes by the convective plasma motions, and the 
maximum field strength rises from its initial value of $200~\mathrm{G}$ to a few kilogauss in the intergranular lanes. 
The sequence we obtained is long enough for the relaxation of the model, since the time needed for the magnetic field
redistribution is about 0.1-0.2 hours \citep{voegler1, voegler2}.

A three-dimensional rendering of the velocity and magnetic field is shown in Fig.~\ref{fig0}. Stream lines (red curves) reveal a 
large amount of vortex motions in the upper photosphere. These vortices appear primarily in the intergranular lanes and coincide 
with regions of strong magnetic fields and downflows. 

\begin{figure*}
\includegraphics{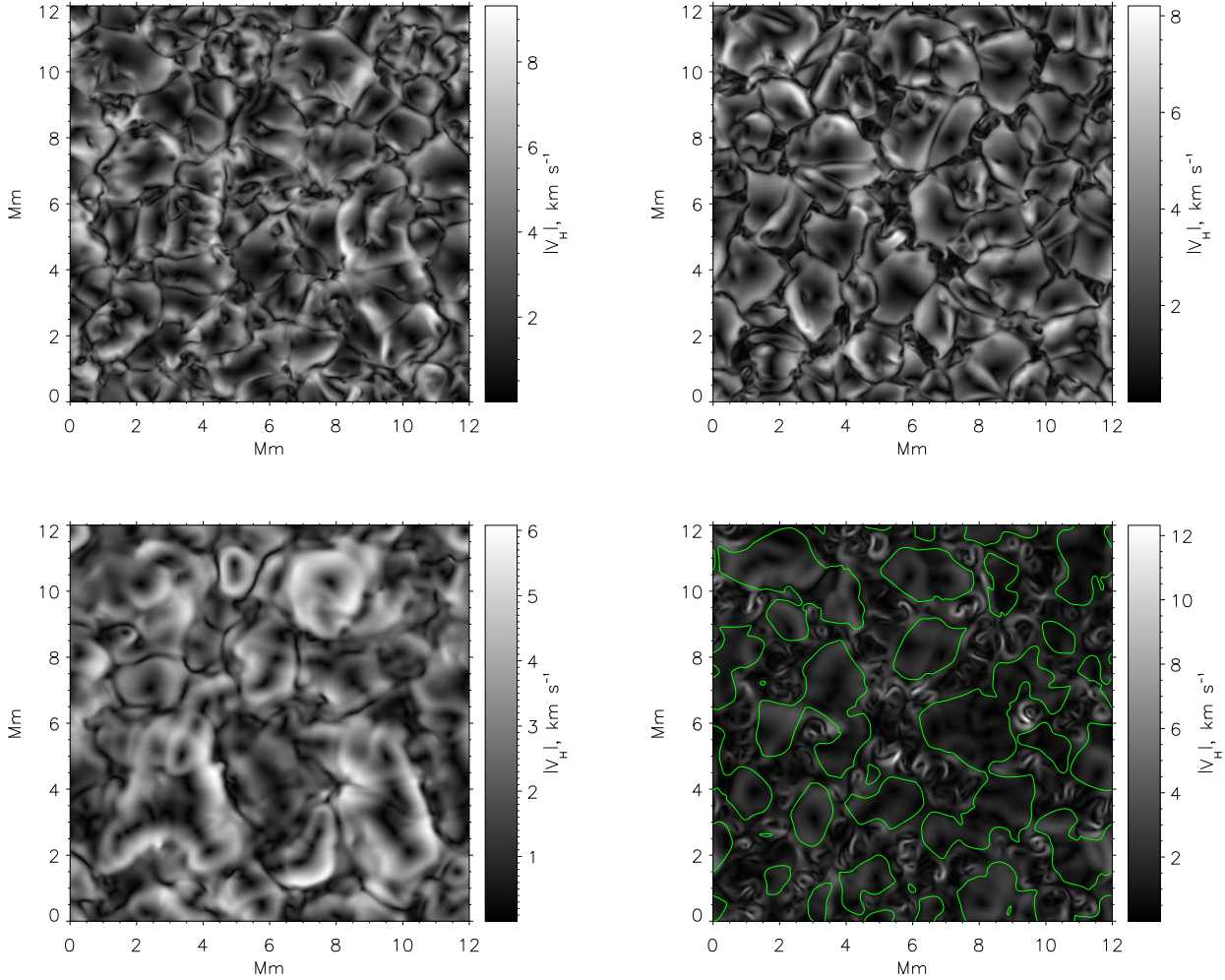}
\caption{Horizontal cuts of the modulus of the horizontal components of velocity in the domain. The cuts are
taken approximately at the visible solar surface (top panels) and in the upper photosphere (bottom panels).
The left panels correspond to the initial non-magnetic snapshot, while the right panels correspond to the well-developed
magnetic model. Vortex motions are clearly visible in the bottom-right panel. The contours on the bottom-right panel
bound the regions where the vertical component of magnetic field $B_y < 30~\mathrm{G}$.}
\label{fig1}
\end{figure*}

The modulus of the horizontal velocity components is shown in Fig.~\ref{fig1}, where the top panels correspond 
to a level close to the visible solar surface level, and the bottom ones to a height in the domain close to the 
temperature minimum. Images on the left correspond to the initial non-magnetic model, and those on the right to the fully 
developed magnetic snapshot. It is evident from the figure that small-scale vortex structures have formed in the
magnetised model in the upper photosphere (bottom-right panel). These structures are not seen in the non-magnetic 
model nor at the visible solar surface. The contours in the bottom-right panel of Fig.~\ref{fig1}, which bound the granular regions 
where the vertical component of magnetic field $B_y < 30~\mathrm{G}$, clearly demonstrate that the vortices 
in the upper photosphere are co-spatial with the magnetic field concentrations in the intergranular network.

\section{The vorticity equation}

\begin{figure*}
\includegraphics{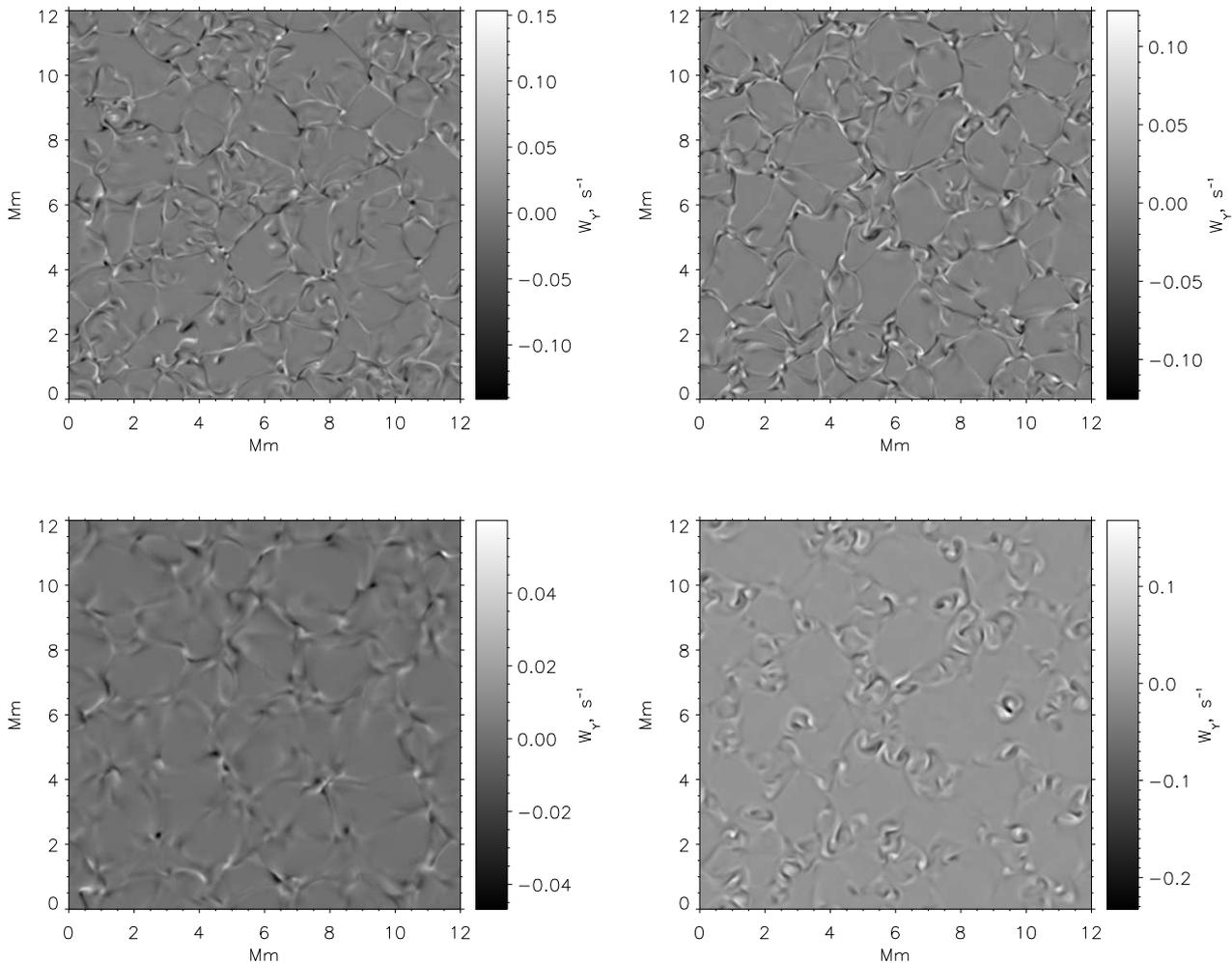}
\caption{Vertical component of vorticity $\omega_y$ at the visible solar surface (upper panels) and in the upper photosphere (lower panels). 
Non-magnetic (left) and magnetic (right) snapshots are shown.}
\label{fig3}
\end{figure*}

Vorticity is defined as the curl of velocity, $\boldsymbol{\omega}=\nabla \times \bf{v}$. Similar to \citet{stein1,moreno1,emonet1,emonet2}, we write the 
vorticity equation as the curl of the momentum equation of the MHD system:

\begin{equation}
\begin{split}
{\rho}\frac{D}{Dt}\frac{\boldsymbol{\omega}}{\rho}=\left(\boldsymbol{\omega} \cdot \nabla\right) {\bf v} - \nabla \frac{1}{\rho} \times \nabla p_{g} + \\
+ \nabla \times \left[ \frac{1}{\rho} \bf{J} \times \bf {B} \right].
\end{split}
\label{voreq1}
\end{equation}

By substituting the current vector $\bf{J}$ with $\nabla \times \bf{B}$ and combining the magnetic pressure term with the gas pressure, we 
obtain the following equation: 

\begin{equation}
\begin{split}
{\rho}\frac{D}{Dt}\frac{\boldsymbol{\omega}}{\rho}=\left(\boldsymbol{\omega} \cdot \nabla\right) {\bf v} - \nabla \frac{1}{\rho} \times \nabla\left(p_{g}+p_{m}\right)  + \\
+ \nabla \times \frac{1}{\rho}\left({\bf B} \cdot \nabla \right) {\bf B},
\end{split}
\label{voreq1a}
\end{equation}

where $D/Dt$ represents the full (material) derivative, $\boldsymbol{\omega}$ is the vorticity vector, $\rho$ and $p_g$ are the plasma 
density and pressure, ${\bf v}$ is the velocity vector, $\bf{B}$ is the magnetic field vector, and $p_m={\bf B}^2/2$ is the magnetic 
pressure. Note that the magnetic field is normalised by a factor $\sqrt{4\pi}$ for convenience. Viscous dissipation of vorticity has been 
neglected in this equation. The material derivative in the left side of Eq.~(\ref{voreq1}) is expressed in terms of Euler derivatives as:

\begin{equation}
{\rho}\frac{D}{Dt}\frac{\boldsymbol{\omega}}{\rho}=\frac{\partial \boldsymbol{\omega}}{\partial t}+{\boldsymbol{\omega}} \left( \nabla \cdot {\bf v} \right) + \left({\bf v} \cdot \nabla \right) {\boldsymbol{\omega}}.
\label{voreq2}
\end{equation}

Eq.~(\ref{voreq1a}) can be rewritten by separating the magnetic and non-magnetic terms and decomposing the last
magnetic term into two:

\begin{equation}
\begin{split}
{\rho}\frac{D}{Dt}\frac{\boldsymbol{\omega}}{\rho}=\overbrace{\left(\boldsymbol{\omega} \cdot \nabla\right) {\bf v}}^{T_1} -\overbrace{ \nabla \frac{1}{\rho} \times \nabla p_{g}}^{T_2} - \\
- \overbrace{\nabla \frac{1}{\rho} \times \left[ \nabla p_m - \left({\bf B} \cdot \nabla \right) {\bf B}\right]}^{T_3} + \overbrace{\frac{1}{\rho}\nabla\times\left[\left({\bf B} \cdot\nabla\right){\bf B}\right]}^{T_4}.
\end{split}
\label{voreq1b}
\end{equation}

The right-hand side of Eq.~(\ref{voreq1b}) shows the different physical mechanisms associated with the generation of vorticity. $T_1$ is the 
vortex tilting term, $T_2$ is responsible for the hydrodynamic baroclinic vorticity generation,  $T_3$ represents the magnetic 
baroclinic vorticity, and $T_4$ corresponds to vorticity generated by the magnetic tension. The quantity in square brackets within $T_3$ represents 
the deviation of the magnetic field configuration from a potential field.

Horizontal cuts of the vertical component of vorticity are shown in Fig.~\ref{fig3}.  The upper panels show the 
vorticity maps at a level close to the visible solar surface, while the lower panels correspond to the upper photosphere. 
Non-magnetic and magnetic snapshots are shown in the left and right panels, respectively.  

In the non-magnetic case, the vorticity is generated by the hydrodynamic baroclinic term and is mostly randomly directed without 
any internal structure of the vortices in the intergranular lanes. However, we notice a change in the structure of the vortices after 
we incorporated the magnetic field into the numerical box. Once the magnetic field has been advected into the intergranular 
network, both positive and negative polarities of the vorticity coexist within a vortex, with both clockwise and counterclockwise 
motions appearing. An increase in the amplitude of the vertical component of vorticity by a factor of about 5 is also observed. 

In order to determine the origin of the vorticity, the terms in Eq.~(\ref{voreq1b}) need to be analysed separately for the whole 
simulation sequence. The vertical component of vorticity, which corresponds to the horizontal vortex motions, is of primary interest. 
Thus, a mean value of the modulus of the vertical vorticity component is a good measure of the amount of vertical vorticity generated in the model.

\begin{figure*}
\includegraphics[width=1.0\linewidth]{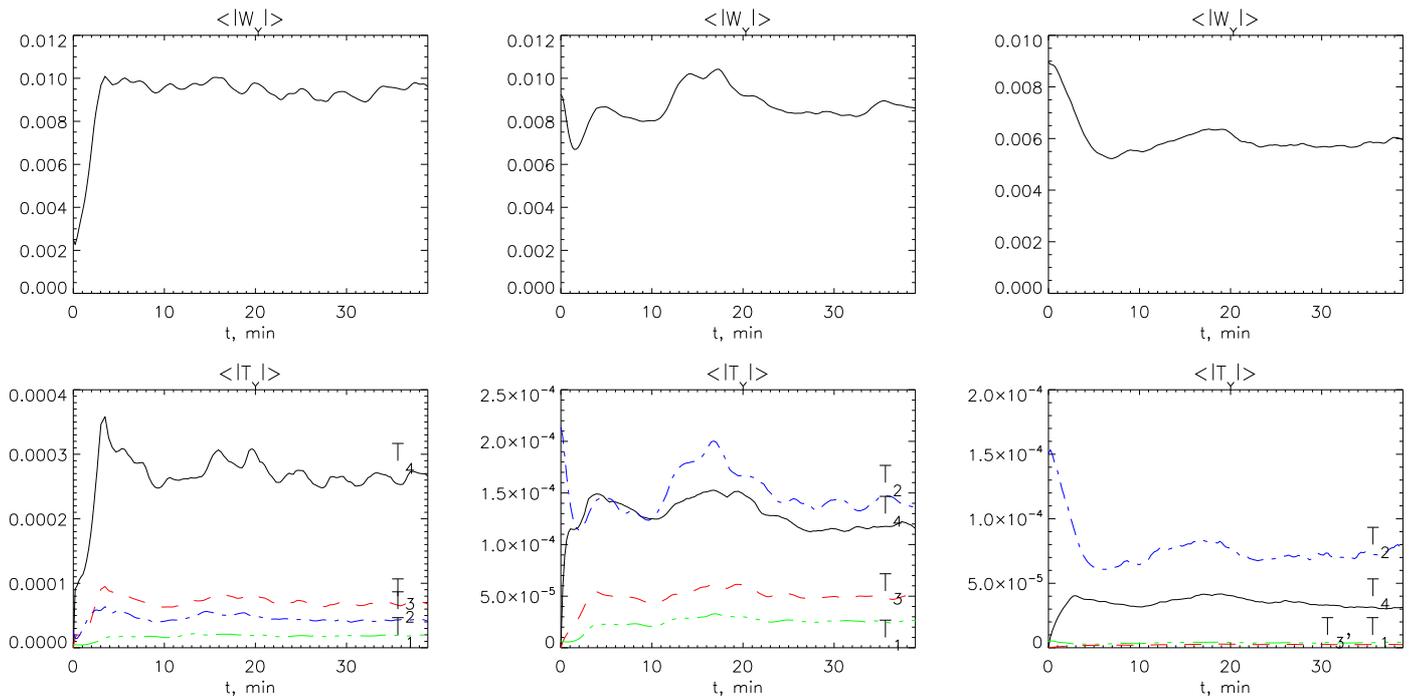}
\caption{Dependence of $<|\omega_y|>$ (top plots) and of different terms of the vorticity equation (bottom plots)
on time at different heights in the domain. The first column corresponds to the upper photosphere ($500~\mathrm{km}$ above the approximate 
visible solar surface level), the second column is the visible surface level, and the third column corresponds to the convection zone ($650~\mathrm{km}$ 
below the visible surface level).}
\label{fig5}
\end{figure*}

Fig.~\ref{fig5} shows the evolution of the mean of the modulus of the vertical vorticity component at different heights
in the domain, together with the time dependences of the $y$-components of $T_1$ through $T_4$ terms of Eq.~(\ref{voreq1b}). The left panels 
correspond to the upper photosphere ($500~\mathrm{km}$ above the visible solar surface), the middle panels 
the approximate level of the visible solar surface, and the right panels correspond to the convection zone 
($650~\mathrm{km}$ below the visible solar surface). 

An inspection of the left panels of Fig.~\ref{fig5} reveals that 
the vorticity in the upper layers of the model is produced by the magnetic field, rising from its non-magnetic value of 
$0.0025~\mathrm{s^{-1}}$  to $0.01~\mathrm{s^{-1}}$ within the first three minutes of the simulation. During this 
phase, the magnetic field gets almost completely transported into the intergranular lanes by convective motions of plasma \citep{voegler1}. The amount 
of vorticity produced in the photosphere (middle-top plot) remains roughly the same, experiencing some initial decrease, which
may be connected to the suppression of plasma motions by the initially uniform magnetic field, before it increases again 
at $t=2-4~\mathrm{min}$. The behaviour of the vorticity at the bottom of the domain is opposite to what is observed at
the top: in the first 5 minutes of the simulation, the amount of vorticity has decreased by almost a factor of 2.

The evolution of vorticity shown in Fig.~\ref{fig5} can be explained by analysing the relative importance of the vertical 
components of the $T_1$ through $T_4$ terms in Eq.~(\ref{voreq1b}). 
In the upper photosphere, the term which corresponds to the vorticity generation by the magnetic tension ($T_4$, solid 
black line) experiences a sharp rise, and after the initial phase of the simulation ($4~\mathrm{min}$) takes the largest value 
among the other terms. The same behaviour is observed for the magnetic baroclinic term $T_3$ (red dashed line), 
with an amplitude which is a factor of 3 smaller than that of $T_4$. The hydrodynamic baroclinic term $T_2$ 
(blue dash-dotted line) is a factor of 2 smaller than $T_3$. Thus, the baroclinic motions of the fluid (in the hydrodynamical sense) do 
not make a significant impact on the vortex generation in the upper photosphere.

A different picture is observed at the photospheric level. Here, the amount of vorticity produced by the hydrodynamic baroclinic motions of the fluid 
is similar to the amount of vorticity, generated by the magnetic tension. In the initial stage of the simulation, $T_4$ and $T_2$ behave in an opposite way. 
This is caused by the processes of magnetic field redistribution. Initially, the magnetic field is uniform, and suppresses the
plasma motions both in the granules and in the intergranular lanes, thus decreasing the vorticity generation by hydrodynamic baroclinic 
term $T_2$, while $T_4$ experiences a sharp increase due to the formation of the magnetic flux concentrations. 
After the initial stage, when the convection pushes the magnetic field out of the granules, both the magnetic-type vortices 
and hydrodynamic vortices can be produced.

In the convection zone, (right panels in Fig.~\ref{fig5}), where the plasma $\beta$ is high, the hydrodynamic baroclinic term $T_2$ is the 
primary source of vorticity. Although $T_2$ decreases significantly from its non-magnetic value within the first few minutes of the simulation, 
it remains a factor of 2 larger than the magnetic tension term $T_4$.

We note that $T_4$ plays a significant role for the vorticity generation both in the convection zone, where the plasma $\beta$ is large, 
and in the upper photosphere, where plasma $\beta$ is small in the magnetic field concentrations. This term has no "hydrodynamic" equivalent
and represents a physical mechanism for vorticity generation which is separate in nature from the conventional baroclinic vorticity 
generation processes. The magnetic baroclinic term $T_3$ does not show any significant influence on the generation of vorticity. 
The significance of the hydrodynamic baroclinic vorticity generation term increases with depth.

$3~\mathrm{mHz}$ acoustic oscillations are also present in the domain. Both the vertical component of vorticity $\omega_y$
and $T_4$ show signs of these oscillations (see upper left plot of Fig.~\ref{fig5}) after the simulation has passed its initial stage.
This finding, together with the connection of the photospheric vortices to the magnetic field, confirms the idea that the oscillations 
leak through magnetic concentrations to the upper photosphere. An observational search for oscillatory signals in vorticity may be possible.

\section{Radiative diagnostics}

\begin{figure*}
\includegraphics[width=1.0\linewidth]{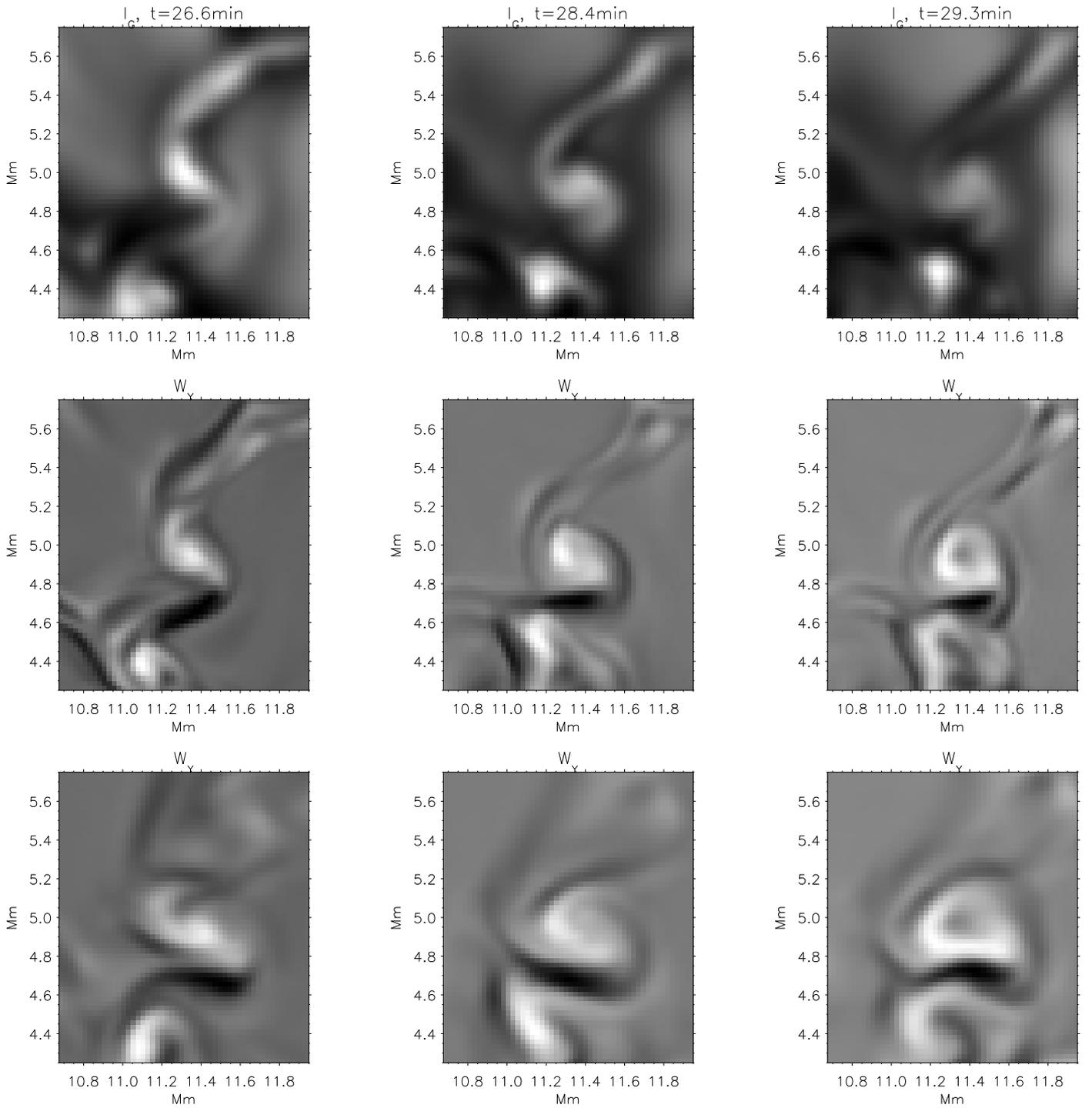}
\caption{Simulated G-band images of an MBP (top panels), vorticity at the visible solar surface (middle panels), and the vorticity in the upper 
photosphere (bottom panels).}
\label{fig6}
\end{figure*}

Photospheric bright points correspond to regions of strong intergranular magnetic fields \citep{shelyagbp1,carlsson,shelyagbp2} and may be subject
to vortex motions. Recent high spatial resolution observations indicate that this may indeed be the case (\citet{bonet1}). The analysis we 
present includes the radiative diagnostics in the G-band. Using the methods described by \citet{frutiger}, \citet{berdyugina} and 
\citet{shelyagbp2}, we have computed 
G-band images for all 147 sequential simulation snapshots. The direct effect of the magnetic field on the absorption line profiles has not been 
included in the calculations. The sequence of images allowed us to not only to find and track the vortex motions of the photospheric G-band 
bright points, but also to study their appearance during the initial stages of the simulation while the magnetic field was being advected to the 
intergranular lanes.  
An inspection of the images made possible to identify vortex motions associated with magnetic bright points (MBPs) in the simulated photosphere. 
An example is shown in Fig.~\ref{fig6} and confirms the connection between the photospheric vortex motions, rotation of MBPs and vorticity in the 
upper layers of the simulated photosphere. Three consecutive simulated G-band images of an MBP (top panels), together with vorticity maps at a 
height corresponding to the visible solar surface (middle) and in the upper photosphere (bottom) are shown in the figure. All images are taken at 
the same horizontal position. The evolutionary track of the MBP is very similar to the observations of \citet{bonet1}. The bottom images clearly 
resemble the evolution of chromospheric swirl \citep{wedemeyer1}, despite the data being obtained somewhat lower than the level of Ca\,{\sc ii} 
core formation. This fact suggests that the chromospheric swirls may be connected to the vortices, generated in the solar magnetic photosphere,
however, a further investigation is needed to provide a rigorous proof for that.

\section{Conclusions}

In this paper, we have examined the processes leading to vorticity generation in the simulated magnetised photosphere.
We have shown that large amount of vorticity in the photosphere is formed as a result of the photospheric plasma interaction with the
magnetic field in the intergranular lanes. The amount of vorticity generated due to the magnetic field is a factor of 4 
larger than that generated by baroclinic motions in convecting photospheric non-magnetic plasma.
By the appropriate decomposition of the magneto-hydrodynamic vorticity equation we defined two vorticity equation terms, which 
are connected to the magnetic field: the first resembles the baroclinic term in hydrodynamics, while the second contains
the magnetic tension and does not have a direct hydrodynamic equivalent. We have demonstrated that it is only the latter term which
is mostly responsible for the generation of vorticity in the upper photosphere. Conversely, the importance of the baroclinic hydrodynamic term 
increases with increasing geometrical depth. We also note the appearance of oscillations in the magnetic tension term in the
upper photosphere. These may be signatures of the $3~\mathrm{mHz}$ lower-photospheric oscillations which leak to the 
upper layers through the intergranular magnetic field concentrations.

Using radiative diagnostics with the G band, we confirmed that MBPs are subject to rotary motions in the intergranular lanes 
and are magnetically connected to the vortices in the upper photosphere.

The large volume occupied by the intergranular vortices suggests their significance for the energy balance of the solar atmosphere. 
Further analysis is needed on the connection of the photospheric vortices with the chromospheric swirls, with the upper chromosphere 
and corona regions, and on torsional wave excitation in these regions of the solar atmosphere.

Follow-up investigations will focus on the details of the physical mechanism, which leads to the creation of negative and positive 
vorticity signs in a magnetic vortex. Extending the simulation box to include the chromosphere with a full non-LTE treatment of the 
radiative diagnostics \citep{carlsson2}, will also be the subject of a future investigation.

\section*{Acknowledgements}
This work has been supported by the UK Science and Technology Facilities Council (STFC). 
FPK is grateful to AWE Aldermaston for the award of a William Penney Fellowship.


\begin{thebibliography}{}

\bibitem[Berdyugina et al.(2003)]{berdyugina} Berdyugina, S.~V., Solanki, S.~K., \& Frutiger, C.\ 2003, \aap, 412, 513 

\bibitem[Bonet et al.(2008)]{bonet1} Bonet, J.~A., M{\'a}rquez, I., S{\'a}nchez Almeida, J., Cabello, I., \& Domingo, V.\ 2008, \apjl, 687, L131 

\bibitem[Carlsson et al.(2004)]{carlsson} Carlsson, M., Stein, R.~F., Nordlund, {\AA}., \& Scharmer, G.~B.\ 2004, \apjl, 610, L137 

\bibitem[Carlsson et al.(2010)]{carlsson2} Carlsson, M., Hansteen, V.~H., \& Gudiksen, B.~V.\ 2010, arXiv:1001.1546 

\bibitem[Cheung et al.(2008)]{cheung1} Cheung, M.~C.~M., Sch{\"u}ssler, M., Tarbell, T.~D., \& Title, A.~M.\ 2008, \apj, 687, 1373 

\bibitem[Clyne et al.(2007)]{vapor} Clyne, J., Mininni, P., Norton, A., \& Rast, M.\ 2007, New Journal of Physics, 9, 301 

\bibitem[Danilovic et al.(2010)]{danilovic2} Danilovic, S., Sch{\"u}ssler, M., \& Solanki, S.~K.\ 2010, \aap, 509, A76 

\bibitem[Danilovic et al.(2010)]{danilovic1} Danilovic, S., Sch{\"u}ssler, M., \& Solanki, S.~K.\ 2010, \aap, 513, A1 

\bibitem[Emonet \& Moreno-Insertis(1998)]{emonet1} Emonet, T., \& Moreno-Insertis, F.\ 1998, \apj, 492, 804 

\bibitem[Emonet et al.(2001)]{emonet2} Emonet, T., Moreno-Insertis, F., \& Rast, M.~P.\ 2001, \apj, 549, 1212

\bibitem[Fedun et al.(2009)]{fedun1} Fedun, V., Erd{\'e}lyi, R., \& Shelyag, S.\ 2009, \solphys, 258, 219 

\bibitem[Frutiger(2000)]{frutiger} Frutiger, C.\ 2000, Ph.D.~Thesis No13896, ETH Z{\"u}rich

\bibitem[Gruszecki et al.(2010)]{nakariakov1} Gruszecki, M., Nakariakov, V.~M., van Doorsselaere, T., \& Arber, T.~D.\ 2010, Physical Review Letters, 105, 055004 

\bibitem[Jess et al.(2009)]{Jess1} Jess, D.B., Mathioudakis, M., Erd{\'e}lyi, R., Crockett, P.J., Keenan, F.P., Christian, D.J., Science, 323, 1582, 2009.

\bibitem[Kitiashvili et al.(2010)]{kosovichev1} Kitiashvili, I.~N., Kosovichev, A.~G., Wray, A.~A., \& Mansour, N.~N.\ 2010, \apj, 719, 307 

\bibitem[Moreno-Insertis \& Emonet(1996)]{moreno1} Moreno-Insertis, F., \& Emonet, T.\ 1996, \apjl, 472, L53 

\bibitem[Parker(1988)]{parker1} Parker, E.~N.\ 1988, \apj, 330, 474 

\bibitem[Pietarila Graham et al.(2009)]{pietarila1} Pietarila Graham, J., Danilovic, S., \& Sch{\"u}ssler, M.\ 2009, \apj, 693, 1728 

\bibitem[Rempel et al.(2009)]{rempel1} Rempel, M., Sch{\"u}ssler, M., Cameron, R.~H., \& Kn{\"o}lker, M.\ 2009, Science, 325, 171 

\bibitem[Sch{\"u}ssler et al.(2003)]{shelyagbp1} Sch{\"u}ssler, M., Shelyag, S., Berdyugina, S., V{\"o}gler, A., \& Solanki, S.~K.\ 2003, \apjl, 597, L173

\bibitem[Sch{\"u}ssler \& Rempel(2005)]{schremp} Sch{\"u}ssler, M., \& Rempel, M.\ 2005, \aap, 441, 337 

\bibitem[Shelyag et al.(2004)]{shelyagbp2} Shelyag, S., Sch{\"u}ssler, M., Solanki, S.~K., Berdyugina, S.~V.,  V{\"o}gler, A.\ 2004, \aap, 427, 335 

\bibitem[Shelyag et al.(2007)]{shelyag2} Shelyag, S., Sch{\"u}ssler, M., Solanki, S.~K., V{\"o}gler, A.\ 2007, \aap, 469, 731 

\bibitem[Solanki(1987)]{solanki} Solanki, S.~K.\ 1987, Ph.D.~Thesis No8309, ETH Z{\"u}rich

\bibitem[Stein \& Nordlund(1998)]{stein1} Stein, R.~F., \& Nordlund, A.\ 1998, \apj, 499, 914 

\bibitem[V{\"o}gler et al.(2005)]{voegler1} V{\"o}gler, A., Shelyag, S., Sch{\"u}ssler, M., Cattaneo, F., Emonet, T., \& Linde, T.\ 2005, \aap, 429, 335 

\bibitem[V{\"o}gler \& Sch{\"u}ssler(2007)]{voegler2} V{\"o}gler, A., \& Sch{\"u}ssler, M.\ 2007, \aap, 465, L43 

\bibitem[Yelles Chaouche et al.(2009)]{yelles1} Yelles Chaouche, L., Cheung, M.~C.~M., Solanki, S.~K., Sch{\"u}ssler, M., \& Lagg, A.\ 2009, \aap, 507, L53 

\bibitem[Wedemeyer-B{\"o}hm \& Rouppe van der Voort(2009)]{wedemeyer1} Wedemeyer-B{\"o}hm, S., \& Rouppe van der Voort, L.\ 2009, \aap, 507, L9 

\bibitem[Wedemeyer-B{\"o}hm(2009)]{wedemeyer2} Wedemeyer-B{\"o}hm, S.\ 2009, arXiv:0911.5639 


\end{thebibliography}
\end{document}